\documentstyle[12pt,amsmath,cite]{article}
\newcommand{\ep}{\varepsilon}
\newcommand{\si}{\sigma}
\newcommand{\de}{\delta}
\newcommand{\al}{\alpha}
\renewcommand{\t}{\theta}
\newcommand{\om}{\omega}
\newcommand{\Om}{\Omega}
\newcommand{\De}{\Delta}
\newcommand{\arctg}{\mbox{arctg}}
\begin{document}
\begin{center}
{\Large\bf
Search\strut{} for periodicities in experimental\strut{} data using
an autoregression\strut{} data model\strut}
\vskip 5mm
B.Z. Belashev${^{1)}}$ and M.K. Suleymanov${^{2,3)}}$
\vskip 5mm

{\small ${^{1)}}$ Institute of Geology, Karelian Research Centre, RAS, Petrozavodsk, Russia\\
${^{2)}}$ LHE JINR, Dubna, Moscow region, Russia\\
${^{3)}}$ NPI ASCR, Rez near Prague, Czech Republic}
\end{center}
\vskip 5mm
\centerline{\bf Abstract}

To process data obtained during interference experiments in high-energy 
physics, methods of spectral analysis are employed. Methods of spectral 
analysis, in which an autoregression model of experimental data is used, 
such as the maximum entropy technique as well as Pisarenko and Prony's  
method, are described. To show the potentials of the methods, experimental 
and simulated hummed data are discussed as an example.
\medskip
 
An alternative for an increase in the energy of colliding particles,
observed when studying the properties of a substance at short distances,
is the observation and study of the f\/ine ef\/fects of the bound states of 
atoms and elementary particles~\cite{[1]}. Of special interest are the interference 
energies and phases of states sensitive to the behaviour of the potential 
at short distances and to external ef\/fects~\cite{[2]}. Information on the bound 
states of systems is obtained in interference experiments~\cite{[3]}. The data 
obtained during such an experiment are represented by oscillating relations 
the spectral parameters of which are assessed to cast light on the energy and 
phase characteristics of a system~\cite{[4]}. The reliability of the resultant 
information depends on the ability to distinguish periodicities in experimental 
data and to estimate their parameters. 

The mathematical theory and algorithmic apparatus, used to reveal latent 
periodicities, have an extensive background and various practical applications. 
Methods for identif\/ication of periodicities are subdivided into a) 
approximation methods that enable a person to approximate experimental 
data by a function which a priori agrees with the preset model and b) 
f\/iltration methods that provide information on components and their 
parameters when a priori evidence is minimum. The spectral characteristics 
of a signal can be estimated more accurately by approximation methods than 
by f\/iltration methods~\cite{[5]}.

The authors discuss a search for latent periodicities in dif\/ferently generated 
data using methods based on an autoregression data model (ARDM)~\cite{[6]}.

ARDM-based methods have an intermediate position between the groups of data 
processing methods discussed. In these methods, approximation is represented 
by a postulate of local data connection and f\/iltration by operations done 
to produce a spectrum and to identify components. Owing to their intermediate 
position, ARDM methods take the advantages of each of the groups described. 
Spectral parameters and resolution can be estimated more accurately by 
ARDM methods than by f\/iltration methods on the basis of FourierÒs discrete 
transformation (FDT)~\cite{[7],[8]}.

Unlike FDT algorithms in which the signal observed is split into harmonic 
components, some of ARDM methods (e.g. Prony's method) use decomposition 
into components in the form of damping oscillations. When processing 
interference experimental data, such a representation has been found 
to be more realistic than a conventional one because it considers the 
lifetime of states. This tendency persisted in Wavelet analysis, where 
decomposition into soliton-like components made it possible to follow 
space-time variations in the spectral characteristics of signals~\cite{[9]}.

Oscillation phases can be estimated more accurately by Prony's method 
than by FDT method. This advantage is important, considering the 
informative capacity of phases in interference experiments. 

ARDM modelling was done in the present study using the spectral maximum 
entropy data technique (MENT), Pisarenko's method and Prony's 
method~\cite{[10],[11],[12]}, simulated and experimental data being used as an example. 

{\bf Autoregression data model}.
ARDM of a random process postulates a local data connection: the current 
value of the function $y(l)$, which approximates the process studied, is 
estimated from the previous data counts $x(l-k)$ in the form of a linear 
combination:
\begin{equation}
y(l) = \sum\limits_{k = 1}^p {a(k)x(l - k)} 
\end{equation}
with depth $p$ and parameters $a(k)$. The parameters of $a(k)$ contain all
information on the spectral characteristics of the process~\cite{[13]}.

Deducting the value of the variable $x(l)$ from both parts of equality (1), 
the error in the current value $\ep(l) = x(l)$ is expressed as:
\begin{equation}
\ep(l) = \sum\limits_{k = 0}^p {a(k)x(l - k)} 
\end{equation}
with the parameter $a(0) = -1$. Because the error in the current value is 
also represented as a linear combination of previous data counts, the 
parameters of $a(k)$ can be regarded as the coef\/f\/icients of a linear f\/ilter 
which predicts an error. 

An error in predicting the current error value can be minimized, for 
example, on the basis of the least squares principle. In the case of an 
independent signal and additive noise, the observed data supplied to the 
input of the error prediction f\/ilter are converted to white noise. It is an 
example of a so-called ``bleaching'' f\/ilter.

Multiplying both parts of relation (2) by the complex conjugate 
value $x^*(l)$ and averaging for all observed data, we will get 
a system of linear equations (3) for the parameters of $a(k)$:
\begin{equation}
\sum_{k=0}^pR(m-k)a(k)=\si^2\de(m)
\end{equation}
where $\si^2$ is noise dispersion, $\de(m)$ is delta-function, $R(m-k)$
is the autocorrelation function values of $R(k)$, $m=0,1,2\dots p$.

To determine the f\/ilter coef\/f\/icients of $a(k)$, Derbu-Levinson's
algorithm~\cite{[14]} is commonly used by consecutively estimating bleaching f\/ilter coef\/f\/icients 
more accurately and calculating the power of the error predicted, beginning 
with a f\/irst-order f\/ilter and ending with $p$-th-order f\/ilter. The algorithm
calculates the sequence
$(a_{11},\sigma_1^2)$, $(a_{21},a_{22},\sigma_2^2),\dots$,
$(a_{p1},a_{p2},\dots,a_{pp},\sigma_p^2)$ using the formulas:
$$
a_{11}=-R(1)/R(0);\quad\sigma_1^2=(1-|a_{11}|^2)R(0)
$$
\begin{equation}
a_{kk}=-\left[R(k)+\sum\limits_{i=1}^{k-1}a_{k-1,i}R(k-i)\right]/\sigma_{k-1}^2
\end{equation}
$$
a_{ki}=a_{k-1,i}+a_{kk}a_{k-1,k-i}^*;\quad\sigma_k^2=(1-|a_{kk}|^2)\sigma_{k-1}^2
$$

The rapidity and economy of Derbu-Levinson's algorithm are due to the fact 
that, unlike the algorithm used to solve systems of Gauss linear equation, 
it requires $p^2$ operations instead of $p^3$ operations and uses the results 
of previous calculations. More ef\/f\/icient algorithms have been developed 
for the same purposes~\cite{[15]}.

{\bf Maximum entropy technique (MENT)}.
As a spectral method, it was f\/irst proposed by Burg~\cite{[10]} to maximize the 
entropy density functional of the spectral capacity of a process $P(l)$
\begin{equation}
-\sum\limits_{l=-(n-1)/2}^{(n-1)/2}\ln P(l)\to\max
\end{equation}
when fulf\/illing the conditions of Winer-Hinchin's theorem for $p+1$ known
values of the autocorrelation function
\begin{equation}
\sum_{l=-(N-1)/2}^{(N-1)/2}P(l)\exp(-j\De\om lk\De t)=R(k),
\end{equation}
where $\De\om$ and $\De t$ are frequency and time intervals between spectrum 
and data counts, $j$ is imaginary unit ($j^2=-1$), $k=0,1,2,\dots p$.

For the Gaussian random process, the requirement of a maximum entropy 
functional is equivalent to the minimum functional of a predicted error. 
This variation problem with the bleaching f\/ilter coef\/f\/icients of $a(k)$ as
Lagrange factors is solved using the formula:
\begin{equation}
P(l)=\frac{\sigma^2}%
{\left|1 + \sum\limits_{k=1}^p a(k)\exp(-j\Delta\omega lk\Delta t)\right|^2}
\end{equation}
which has a simple meaning: the power spectrum of a signal is found by 
dividing the output power of noise by the squared module of the spectral 
characteristics of the bleaching f\/ilter. The coef\/f\/icients of $a(k)$ are 
estimated in ARDM by solving a system of equations (3).

MENT considers local data connection, estimates the power spectrum of 
noise, does not result in negative values in the power spectrum and has 
a better resolution as compared with FDT, but is inferior in resolution 
to Pisarenko's and Prony's methods. As MENT is equivalent to the least 
squares method, it gives an undisplaced estimate of the spectrum. The 
requirement of maximum entropy or minimum information at preset restrictions 
automatically excludes all alien frequency peaks from the estimate of 
the spectrum. For this reason, the spectral peaks in MEM that correspond 
to an anharmonic periodic histogram are expected to be weaker than in FDT~\cite{[16]}.

MENT is ef\/f\/icient for express estimation of the frequency spectrum of random 
processes. It can also be employed to reveal latent periodicities. To show 
this, let us discuss the dependence of the ionization potential of an 
atom in basic state on the atomic number of an element~\cite{[17]} (Fig.1a) and 
the time dependence of the rate of water f\/low through the Solomensky 
Strait of the Petrozavodsk Bay in Lake Onega~\cite{[18]} (Fig. 2a). To produce 
frequency spectra, programme~\cite{[19]} was used in these examples.

The frequency spectrum in Figure 1b shows periods 8, 10, 18 and 32 
characteristic of Mendeleyev's Table. The FDT spectrum of such an anharmonic 
distribution contains, in addition to the main periods $T_{k,l}$, the periods
$\displaystyle T_{k,i}=T_{k,i-1}\frac{i-1}{i}$
equal to 4.0, 2.7 and 2 for the most intensive and narrow peak 8. In this 
spectrum, poorly intensive peaks are far more numerous, and among the 
periods 6.0, 4.7, 4.1, 3.1, 2.8, 2.3, 2.2 and 2.1 only 4.1, 2.8 and 2.1 can 
be considered to be close to anharmonic periods.

In the second example, the pattern of the signal studied is fairly close to 
a combination of harmonics and a noise component. In the power spectrum of 
this signal (Fig. 2b), all medium- and high-frequency peaks were interpreted. 
These peaks were correlated in the order of increasing periods with the seiches 
oscillations of the water body connected through the strait with the 
Petrozavodsk Bay, those of the Petrozavodsk Bay and those of Great Lake Onega. 
Correlation with observed f\/ield data and simulation data gave consistent 
results~\cite{[18]}. Spectral analysis revealed a ca. 12 hour tidal period, 
unknown earlier for lakes, and other low-frequency periods that were hard 
to interpret. The spectral resolution of MEM proved to be quite suf\/f\/icient 
for analysis of the results obtained.

{\bf Pisarenko's and Prony's methods}. Pisarenko's method, used to distinguish 
harmonic components from their combination with white noise, provides even 
higher spectral resolution than MENT. It became possible to increase spectral 
resolution in Pisarenko's method at least 1000 times as compared with that of 
MENT. In this method, the frequencies of components are calculated, whereas 
in FDT or MENT they are estimated visually from the positions of the peaks 
of the frequency spectrum. Frequency estimation accuracy and spectral 
resolution are preset by the accuracy of calculations.

To show the algorithm used to estimate frequencies in Pisarenko's method, 
let us discuss the sinusoid $x(l)=sin(\Om l)$. The trigonometric identity
\begin{equation}
\sin(\Omega l)=2\cos\Omega\cdot\sin(\Omega(l-1))-\sin(\Omega(l-2))
\end{equation}
provides a link between data:
\begin{equation}
x(l)=2\cos\Omega\cdot x(l-1)-x(l-2)
\end{equation}
The Fourier's transformation of relation (9) results in the relation:
\begin{equation}
X(\omega)\cdot\left(1-2\cos\Omega\cdot z^{-1}+z^{-2}\right)=0,
\end{equation}
where z is understood as $\exp(-j\Omega)$. The roots of the equation of 
second degree $z_1=z_2^*$ determine the frequencies $\Om$ and $-\Om$  in 
accordance with the expression
\begin{equation}
\Omega=\arctg(\mbox{Im}z_i/\mbox{Re}z_i).
\end{equation}
In a general case, local data connection in Pisarenko's method is preset 
in the form:
\begin{equation}
x(l)=-\sum\limits_{k=1}^{2p}a(k)y(l-k)+\varepsilon(l),
\end{equation}
and for the coef\/f\/icients of $a(k)$ a system of linear equations:
\begin{equation}
\sum\limits_{k=0}^{2p}R(m - k)a(k)=\sigma^2a(m)
\end{equation}
is obtained.
	
The coef\/f\/icients $a(k)$ and noise dispersion $\si^2$ are estimated by 
calculating the eigen numbers and eigen vectors of the matrix of 
autocorrelation functions $R(m,k)$~\cite{[20]}. Dispersion $\si^2$ corresponds 
to the minimum eigen number of the matrix $R(m, k)$. In this case, it is 
convenient to use the recursive expression
$R\Vec{c}(l + 1)=\Vec{c}(l)$, which follows from (13), and to determine the
vector $\Vec{c}(l + 1)$ from the vector $\Vec{c}(l)$ estimated from 
previous iteration. It is convenient to take the vector $\Vec{c}(0)=[1,1,1,...,1]$
for initial approximation. Several iterations make it possible to produce 
a vector close to $\Vec{c}(\infty)$. It is used to estimate
$\displaystyle\lambda_{\min}=\sigma^2=\frac{\Vec{c}^{\;T}R\Vec{c}}{\Vec{c}^{\;T}\Vec{c}}$
and vector $\Vec{a}$ is determined as $\displaystyle\Vec{a}=\frac{\Vec{c}}{\lambda_{\min}}$.

The structure of the matrix of the autocorrelation functions $R(m, k)$ is 
such that if harmonics are present in the process studied, then the f\/ilter 
coef\/f\/icients $a(k)$ are valid and meet the requirement: $a(k)=a(2p-k)$. 
Composing and solving an algebraic equation of $2p$ degree with the real 
coef\/f\/icients $a(k)$ and $a(0)=1$
\begin{equation}
\sum\limits_{k = 0}^{2p}a(k)z^{2p - k}  = 0,
\end{equation}
complex conjugate roots, equal in module to unit, are calculated. The 
roots determine the frequencies of harmonics in accordance with expression (11).

The positive frequencies of harmonics $\Om_i$, $i=1,2,\dots,p$, are 
used to estimate spectral power frequencies $P(i)$, solving a system of equations
\begin{equation}
\sum\limits_{i=1}^pP(i)\cos(\Om_il\De t)=R(l).
\end{equation}
The self-consistency of the results obtained is checked by estimating noise 
dispersion $\si^2$ and expressing it by expression:
\begin{equation}
\si^2=R(0)-\sum_{k=1}^pP(i).
\end{equation}
Using Pisarenko's method, one can estimate the dispersion of additive white 
noise. However, obtaining non-negative estimates of the spectral density of 
harmonics is not guaranteed. Like in MENT, the initial phase of harmonics 
is not determined in Pisarenko's method.

This shortcoming is overcome in Prony's method in which Pisarenko's idea 
is generalized for non-steady periodic processes, using damping oscillations 
as a basis for spectral decomposition. 

In Prony's method, the function $y(l)$, which approximates the process, is 
expressed through the complex numbers $b_i$ and $z_i$ by the expression
\begin{equation}
y(l) = \sum\limits_{i = 1}^{2p}b_i\cdot z_i,\quad l=0,1,2,...,n-1,
\end{equation}
\begin{align*}
b_i&=A_i\exp(j\theta _i)\\
z_i&=\exp(\alpha_i+j\Omega_i\Delta t)
\end{align*}
where the parameters $A_i$, $\al_i$, $\Om_i$ and $\t_i$ are the amplitude, 
coef\/f\/icient of damping, frequency and initial phase of the $i$-th component 
of the decomposition of reference data.

Local connection between data is postulated like in Pisarenko's method, the 
coef\/f\/icients $a(k)$ are also calculated and used to compose polynomial (14). 
The dif\/ference lies in the fact that the roots of the polynomial are now 
represented by complex exponents $z_i$.

Proceeding from the minimum least squares functional
\begin{equation}
\sum\limits_{l=0}^{n-1}(x(l)-y(l))^2\to\min,
\end{equation}
the vector $\Vec{B}=(b_1,b_2,b_3,\dots,b_{2p})$, which is dependent on the relation
\begin{equation}
\Vec{B}=(F^\# F)^{-1}F^\#\Vec{X},\end{equation}
is estimated. In the above relation, $\Vec{X}$
is a data line, $F$ is Van der Mond's matrix composed of the degrees of roots
$z_i$, $F^\#$ is a transposed matrix. From this relation and the roots of 
polynomial $z_i$ the parameters of decomposition components are estimated:
\begin{align}
& A_i=|b_i|;\quad\Omega_i=\arctg({\mathop{\rm Im}\nolimits}z_i/{\mathop{\rm Re}\nolimits}z_i)\nonumber\\
&\al_i=(\ln|z_i|)/\De t;\quad\t_i=\arctg({\mathop{\rm Im}\nolimits}b_i/{\mathop{\rm Re}\nolimits}b_i)
\end{align}	

Finally, Prony's method gives a data estimate:
\begin{equation}
s(t)=\sum\limits_{i=1}^{2p}A_i\exp(\alpha_i t)\cdot\exp(j(\omega_i t+\theta_i)),
\end{equation}
its Fourier image:
\begin{equation}
S(\omega)=\sum\limits_{i=1}^{2p}A_i\exp(j\theta_i)\frac{1}{(\omega-\omega_i)-j\alpha_i}
\end{equation}
and an energy spectrum:
\begin{equation}
E(\omega)=|S(\omega)|^2.
\end{equation}

It has already been noted that an advantage of PronyÒs method is a broader 
class of functions used as a basis of spectral decomposition. The formalism of 
the method allows the incorporation of exponentially falling and rising signals 
into this class. Therefore, methods of spectral analysis can be applied to 
monotonically varying and transitional random processes. The opportunity to 
calculate the amplitudes, coef\/f\/icient of damping, frequency and especially 
phase of oscillations can be used to interpolate and extrapolate signals. 
In some cases, this helps forecast processes, which is a dif\/f\/icult problem. 

To simulate PronyÒs method, the authors wrote a programme in Fortran. 
Figure 3 shows initial signals in the form of a single sinusoid (a) and 
a mixture of four dif\/ferent harmonics with additive white noise (b) 
and the results of their reconstruction based on the parameters estimated 
by decomposition. For the sinusoid preset by the formula
\begin{equation}
1.600\cos (2.021t + 1.0798)
\end{equation}
with 15\% additive noise the estimate
\begin{equation}
1.594\cos (2.021t + 1.0930)
\end{equation}
was calculated, and for a mixture of harmonics
\begin{align}
& 2.000\cos (0.5054t + 0.2670) + 2.000\cos (1.015t + 0.5422) + \nonumber \\
& + 1.000\cos (1.844t + 0.948) + 1.600\cos (2..465t + 1.3167)
\end{align}
with 6\% noise the estimate 
\begin{align}
& 1.932\cos (0.5054t + 0.2821) + 1.957\cos (1.015t + 0.5887) + \nonumber \\
& + 0.9216\cos (1.844t + 0.9716) + 1.636\cos (2.465t + 1.2530)
\end{align}

The above results are in good agreement with the reference signals. It has 
been shown in these and other computing experiments that Prony's method 
estimates frequencies most accurately and the phases of components and 
the coef\/f\/icients of damping of components least accurately and that the 
spectral parameters of high-frequency components are more accurate than 
those of low-frequency components.

In the ARDM methods, a priori information, preset in the form of local data 
connection, is generalized, can be used to form a model of an oscillating 
signal and to determine its composition and the spectral parameters of 
components without any special a priori assumptions. The resultant parameters 
can be estimated more accurately within the model formed (in some cases, 
it is desirable), using other powerful approximation methods, e.g. the Monte 
Carlo method~\cite{[21]}.

\medskip\noindent     
Conclusions:
\medskip

1. Methods of spectral analysis are highly informative. They are employed 
in interference experiments to cast light on the characteristics of the 
bound state of systems, such as atoms, nuclei and particles. 

2. ARDM-based methods, used to estimate the spectral characteristics of 
signals, combine the advantages of approximation and f\/iltration methods. 

3. The estimates obtained by MENT for the power spectra of real 
signals are informative, and can be used to analyse harmonic signals and 
anharmonic distributions. The spectral parameters of oscillating signals, 
estimated using Prony's method, are highly accurate and allow to represent 
experimental data by the superposition of basic functions.

\listoffigures

Fig. 1. Dependence of the ionization potential of atoms in the main state 
on the atomic number of an element (a) and the MEM estimate of its power 
spectrum (b). The frequency peaks of the power spectrum are arranged in 
accordance with the periods in Mendeleyev's Table.

Fig. 2. Time relation of the current velocity through the Solomensky 
Strait of the Petrozavodsk Bay, Lake Onega (a) and its MEM estimate of 
the power spectrum. Frequency peaks are corresponded by seisches 
oscillation periods of 22--60 min in Lake Logmozero (the lake connected through 
the Solomensky Strait with the Petrozavodsk Bay), 1 hr and 20 minutes 
to 2 hrs and 18 min in the Petrozavodsk Bay, 3 hrs and 8 min to 4 hrs 
and 27 min in Lake Onega and 12 hrs and 24 min in the tidal period.

Fig. 3. The results of the simulation of Prony's method on models: (a) -- a 
simple harmonic signal $1.600\cos(2.021t + 1.078)$ and its estimate 
$1.594\cos(2.021t + 1.0930)$, (b) -- a mixture of harmonics 
$2.000\cos(90.5054t + 0.2670) + 2.000\cos (1.015t + 0.5422) + 1.000%
\cos (1.844t + 0.9848) + 1.6\cos (2.465t + 1.3167)$ and its estimate 
$1.932\cos (0.5054t + 0.2821) + 1.957\cos 91.015t + 0.5887) + 0.9216%
\cos (1.844t + 0.9716) + 1.936\cos (2.465t + 1.2530)$.

\begin{thebibliography}{99}
\bibitem{[1]} I.B. Khriplovich. Non-retention of evenness in atomic phenomena.-M., Nauka, 1981.
\bibitem{[2]} Y.L. Sokolov. Interference method for measuring the parameters of atomic states.- UFN, 199, vol. 69, p. 559-583. 
\bibitem{[3]} M.I. Podgoretsky and O.A. Khrustalev. On some interference phenomena in quantum transitions.- UFN, 1963, vol. 81, p. 217-247.
\bibitem{[4]} E.B. Aleksandrov, G.I. Khvostenko and M.P. Chaika. Interference of atomic states.- M., Nauka, 1991.
\bibitem{[5]} G.I. Vasilenko. Signal regeneration theory.- M., Sov. Radio, 1979.
\bibitem{[6]} S.K. Kay and S.I. Marple. Spectral analysis: a modern prospective. Proc. IEE, vol. 69, n. 11, 1981, p. 1380-1418.
\bibitem{[7]} H. Babic and G.S. Temes. Optimum low-order windows for discrete Fourier transform system. Vol. ASSP-24, Dec. 1976, p. 512-517.
\bibitem{[8]} M.G. Serebrennikov and A.A. Pervozvansky. Revealing latent periodicities. M., Nauka, 1965.
\bibitem{[9]} N.M. Astafyeva. Wavelet-analysis: basic theory and examples of application.- UFN, 1996, vol. 166, p. 1145-1170.
\bibitem{[10]} J.R. Burg. Maximum entropy spectral analysis .- In: Proc. $37^{th}$ Meeting Society of Exploration Geophysicist (Oklahoma City, OK), 31, 1967.
\bibitem{[11]} V.F. Pisarenko. On the estimation of spectra by means of nonlinear functions of a covariance matrix.- Geophysical j. Royal Astronomical Soc., vol. 28, 1972, p. 511-513.
\bibitem{[12]} G.R.B. Prony. Essai experimental et analytique. Paris, J. de L'Ecole Politecnique, vol. 1, cahier 2, 1975, p. 24-76.
\bibitem{[13]} Akaike. Power spectrum estimation through autoregressive model f\/itting.- Ann. Inst. Statist. Math., 1969, vol. 21, p. 243-247.
\bibitem{[14]} B.G. Gubenko. Round of error propagation in Durbin's, Levinson's and Trench's algorithms.- Rec. 1979 IEEE Int. Conf. Acoustic, Speech and Signal Processing, p. 498-501.
\bibitem{[15]} F.B. Hildebrand. Introduction to numerical analysis. Ö New York: McGraw Hill, 1956, ch. 9.
\bibitem{[16]} V.B. Zlokazov. Analysis of latent anharmonic periodicities.- OIYI, P11-89-378, Dubna, 1989.
\bibitem{[17]} M.M. Protodyakonov. The properties and electron structure of rock-forming minerals.- M., Nauka, 1965.
\bibitem{[18]} S.F. Rudnev. Interaction of seisches and discharge current in Lake Onega.- Submitted to VINITI 29.03.89. No. 2060-B89). 19 p.
\bibitem{[19]} I. Barrodale and R.E. Erikson. Algorithms for least-square linear prediction and maximum entropy spectral analysis.- Geophysics, 1980, vol. 45, n. 3, p. 420.
\bibitem{[20]} D.S. Farden. Solution of a Toeplittz set of linear equations.- IEEE Trans. Antennas Propagat., 1976, vol. AP-24, p. 906-908.
\bibitem{[21]} I.M. Sobol. Monte Carlo's numerical methods.- M., Nauka, 1973.
\end{thebibliography}
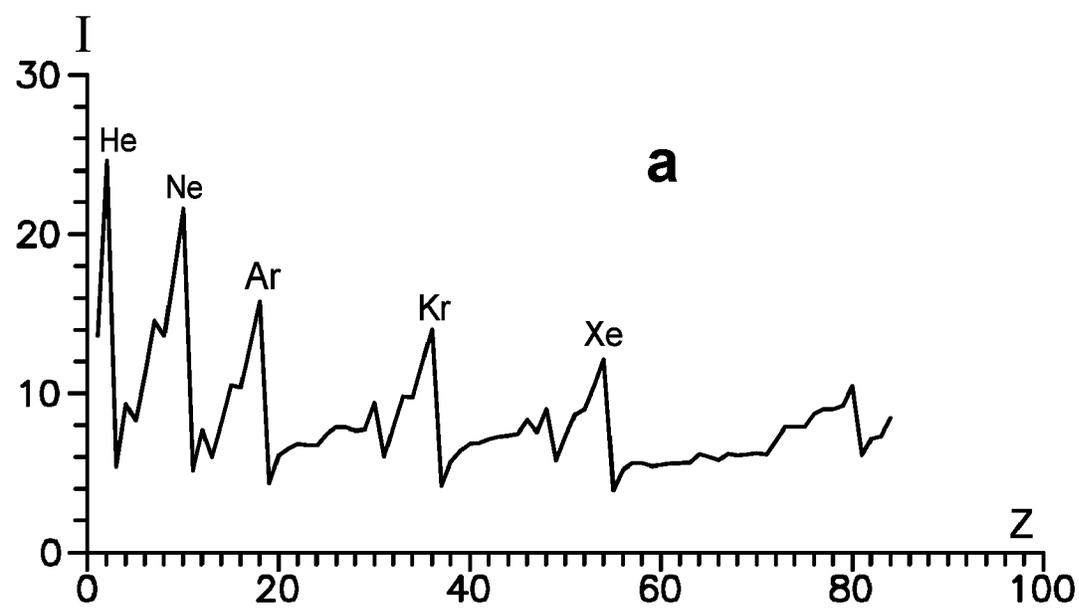
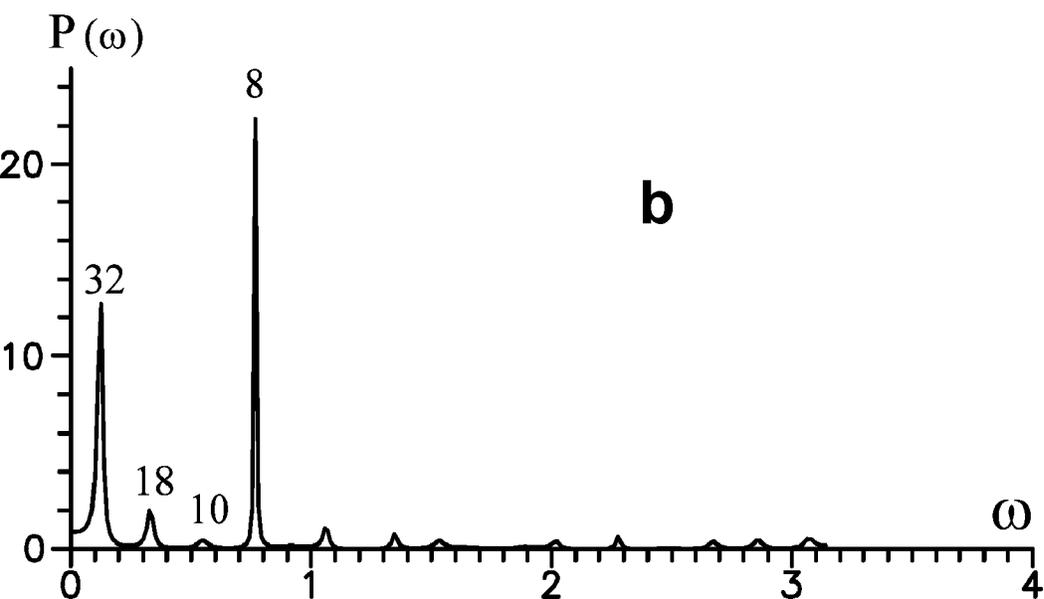
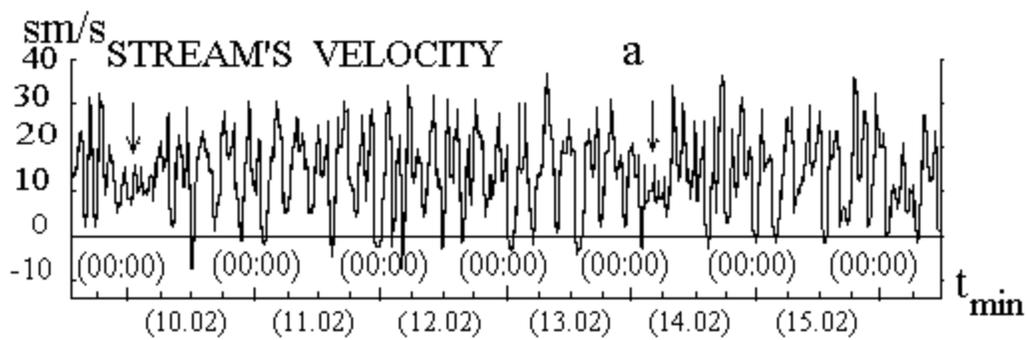
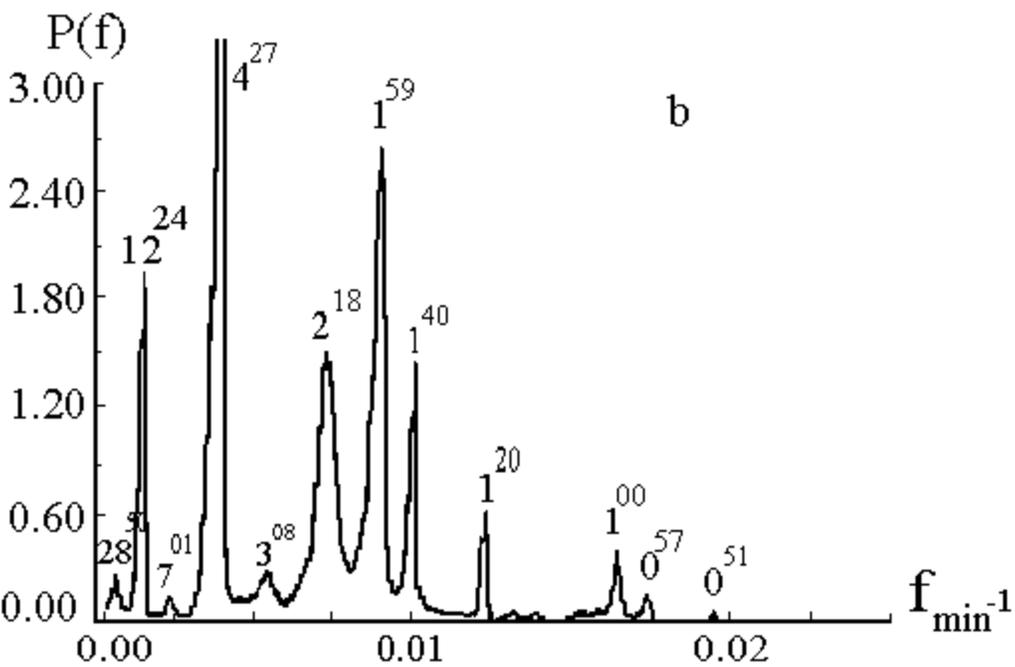
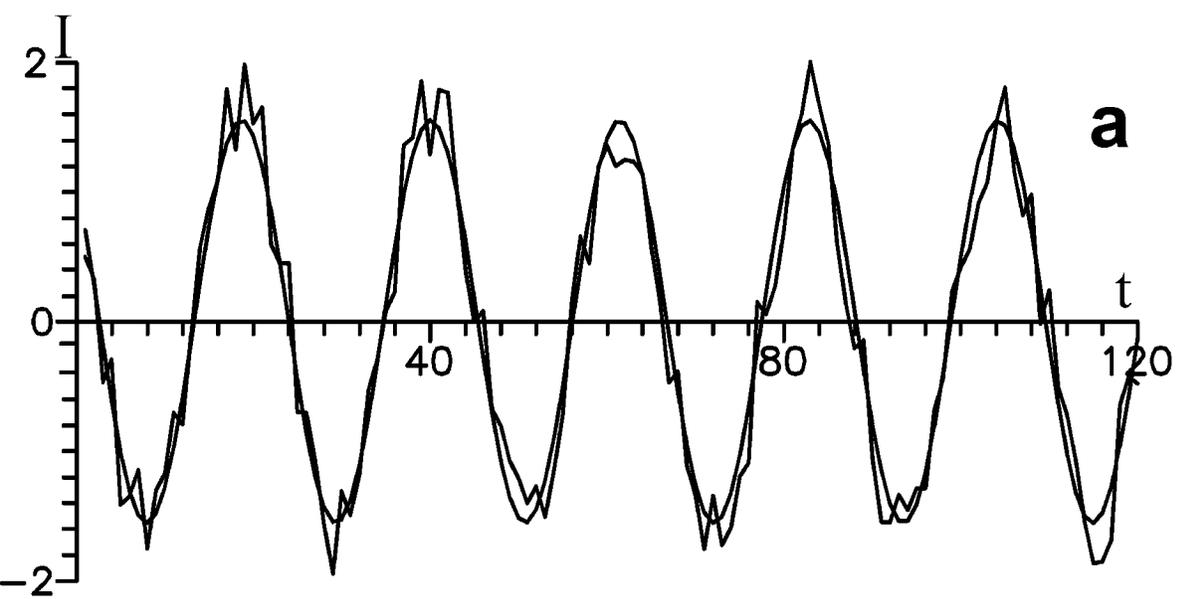
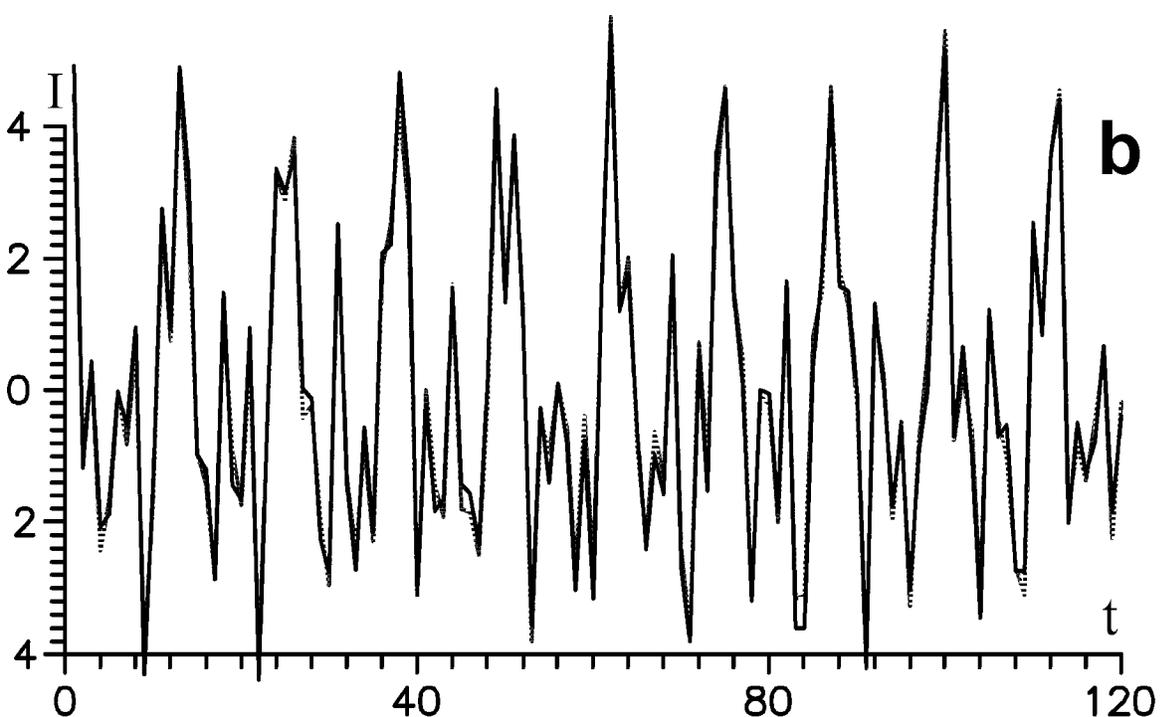
\end{document}